# Fluxonic Cellular Automata

MILORAD V. MILOŠEVIĆ, GOLIBJON R. BERDIYOROV AND FRANÇOIS M. PEETERS*

Departement Fysica, Universiteit Antwerpen (Campus Middelheim), Groenenborgerlaan 171, B-2020 Antwerp, Belgium
*e-mail: Francois.Peeters@ua.ac.be

We formulate a new concept for computing with quantum cellular automata composed of arrays of nanostructured superconducting devices. The logic states are defined by the position of two trapped flux quanta (vortices) in a 2x2 blind-hole-matrix etched on a mesoscopic superconducting square. Such small computational unit-cells are well within reach of current fabrication technology. In an array of unit-cells, the vortex configuration of one cell influences the penetrating flux lines in the neighboring cell through the screening currents. Alternatively, in conjoined cells, the information transfer can be strengthened by the interactions between the supercurrents in adjacent cells. Here we present the functioning logic gates based on this fluxonic cellular automata (FCA), where the logic operations are verified through theoretical simulations performed in the framework of the time-dependent Ginzburg-Landau theory.  The input signals are defined by current loops placed on top of the two diagonal blind holes of the input cell. For given current-polarization, external flux lines are attracted or repelled by the loops, forming the '0' or '1' configuration. The read-out technology may be chosen from a large variety of modern vortex imaging methods, transport and LDOS measurements.

At the time when physical limitations started to foreshadow the eventual end to the traditional scaling in microelectronics, interest has turned to various nanotechnologies vying to succeed traditional CMOS. Quantum dot Cellular Automata (QCA), first described a decade ago by Lent *et al.*[1], is one of these technologies. QCA has proven to be a feasible alternative that could offer higher device densities, faster switching, and lower power. The simplicity of QCA made them well suited for experimental implementation, fabrication and testing[2-4]. However, despite having been physically demonstrated at a small scale, the attractiveness of QCA begins to diminish as logic complexity grows. Leaving this avenue of exploration of novel device architectures, superconducting nanostructured arrays in a magnetic field have been considered for storing information[5]. When a magnetic field is applied to a superconductor, the flux enters in the form of individual quantized vortices which repel each other[6]. In contrast to many other systems, the density of vortices, as well as the strength and range of their interaction, can be easily modified by an external field and temperature. This property makes superconductors ideal for designing micromagnetic flux quanta machines[7-11]. If needed, individual vortices can be trapped by nanoscale dots of non-superconducting material, or by antidots perforating the superconducting films[12-17]. Despite the small size, recent work on confined mesoscopic superconductors[18-21] has demonstrated that individual vortices can be captured in a single sample, singly or multiply connected.

Therefore, in this article we explore the idea of creating superconducting analogue of QCA. As a unit cell, we use a mesoscopic superconducting square, with a regular 2x2 array of blind holes etched from the surface. Unlike antidots, blind holes have a thin superconducting bottom layer, which allows the trapped flux to remain as separated single quantum vortices inside the pinning site[22]. If specific conditions are met, two vortices nucleate in the sample, and their pinning position, as compared to the single electron charges in QCA, determines the logic state. This state can be toggled by fields emerging from the neighboring cells. Since there is no flow of charges of particles and vortices solely change their position in the pinning potential well, an ideal fluxonic cellular automata (FCA) circuit would, in principle, operate near the thermodynamic limit of the information processing. In addition, the interconnect requirements on the nanometer scale are relaxed.

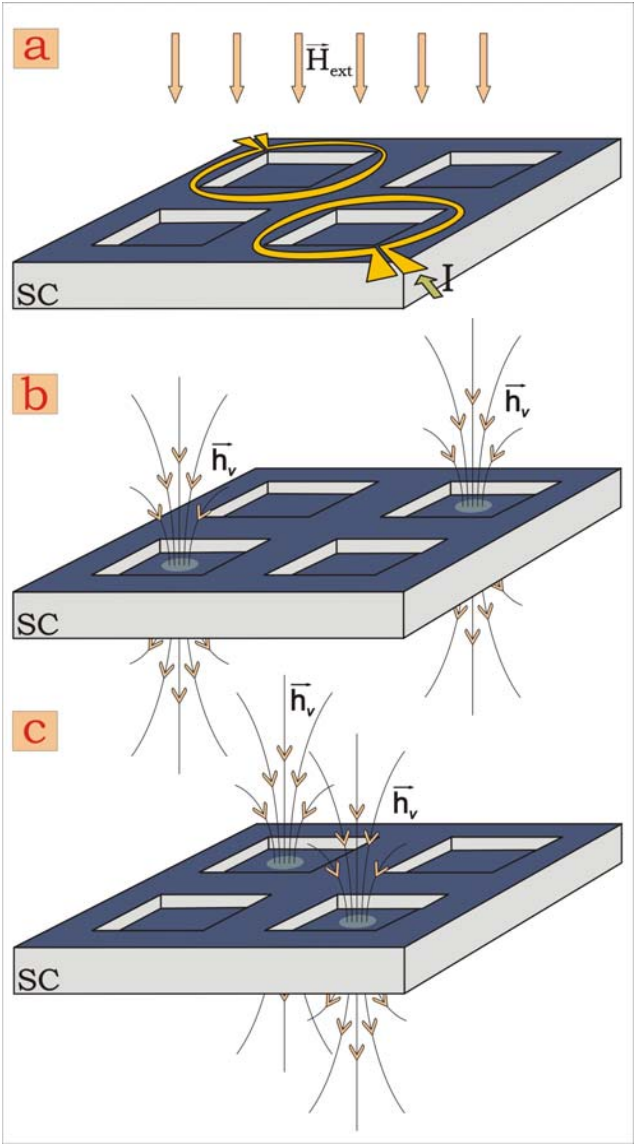

Figure 1 Schematic diagram of the fluxonic cellular automata unit-cell. The superconducting square disk is lithographically etched on the surface in a 2x2 blind holes pattern and exposed to homogenous magnetic field $H_{ext}$. The equilibrium position of the nucleated magnetic vortices is influenced by the current I, circulating in two metallic loops encircling the diagonal blind holes (a). In such a way, two vortices in the sample may switch their position between the two degenerate configurations, determining logic states '1' and '0' [(b) and (c), respectively].

## FCA BASICS

As shown schematically in Figure 1, a unit logic cell of FCA is a square superconducting disk containing a blind hole in each corner. To illustrate the basic FCA phenomena, we consider an aluminum superconducting microsquare with side $w=3.0\mu m$ and thickness $d=150nm$, containing four blind holes with side $w_i=850nm$ and bottom thickness $d_i=30nm$, and an interdistance between the holes of $w_0=350nm$. The operating temperature is taken relatively close to the superconducting/normal state transition, namely $T=0.94T_c$. The coherence length $\xi(0)=120nm$ and the penetration depth $\lambda(0)=140nm$ at zero temperature were estimated for similar Al mesoscopic samples in Ref.[21]. All above given parameters are chosen to suit the standard nano-lithography techniques. Therefore, this example of a FCA unit-cell is easily experimentally realizable, and can serve for educational purposes. Much more sophisticated/smaller devices are needed for actual technological applications, as will be explained later on.

Within the phenomenological Ginzburg-Landau (GL) theory (see Methods), we investigated the superconducting state of this sample, when exposed to a homogeneous external magnetic field $H_{ext}$. It is well known that for superconductors in a magnetic field, in thermodynamic equilibrium the Gibbs free energy reaches its minimum[23]. Therefore, in our search for the steady-state solution of two coupled GL equations for the order parameter $\psi$ and the vector potential $A=rotH$, we compare the energies of all the stable states found. The complete Gibbs free energy ($G$) diagram, as a function of the applied field is shown in Figure 2(a).

As one can see, with increasing applied field $H_{ext}$, the ground state goes through vortex configurations denoted by successive numbers indicating vorticity ($L$). Therefore, external flux lines individually enter the sample, and the vortex states up to vorticity $L=13$ can nucleate, with a superconducting/normal transition field $H_{c3}=4.352mT$. Note that the energy difference between consecutive states is much larger than the thermal activation energy $k_BT_c$, and thermally driven transitions are not possible. However, at temperatures closer to $T_c$, thermal fluctuations may prove to be important, as the saddle-point energies decrease ($G[k_BT_c]\sim(1-T/T_c)^2$). As emphasized in the inset of Figure 2(a), the vortex states show enhanced stability for even vorticity. The physical reason for such behavior follows from the commensurability effects between the vortices and the blind-holes-pattern[12,13,21,24]. The $|\psi|^2$-density contourplots shown in Figure 2(b-e) illustrate the vortex configurations for up to 4 vortices captured in the sample (blue color corresponds to utterly suppressed order parameter, i.e. the vortices).

The vortex state we are particularly interested in is the $L=2$ state, which is found stable (and therefore experimentally realizable) in the applied field range of $H_{ext}=0.2746\div1.4705$ $mT$ ($0.575\div0.824$ $mT$ as the ground-state). When two external flux lines are penetrating the sample, they can form two degenerate ground-state vortex configurations [see Figure 1(b-c)]. Due to their mutual repulsion, the vortices can occupy the blind holes along either one of the diagonals. We use these two states to represent a logic value '1' and a logic value '0' as schematically shown in Figure 1 (b) and (c), respectively.

Therefore, the FCA unit-cell described above now meets all the criteria for building and performing ground state computation. However, the question of predefining the unit-

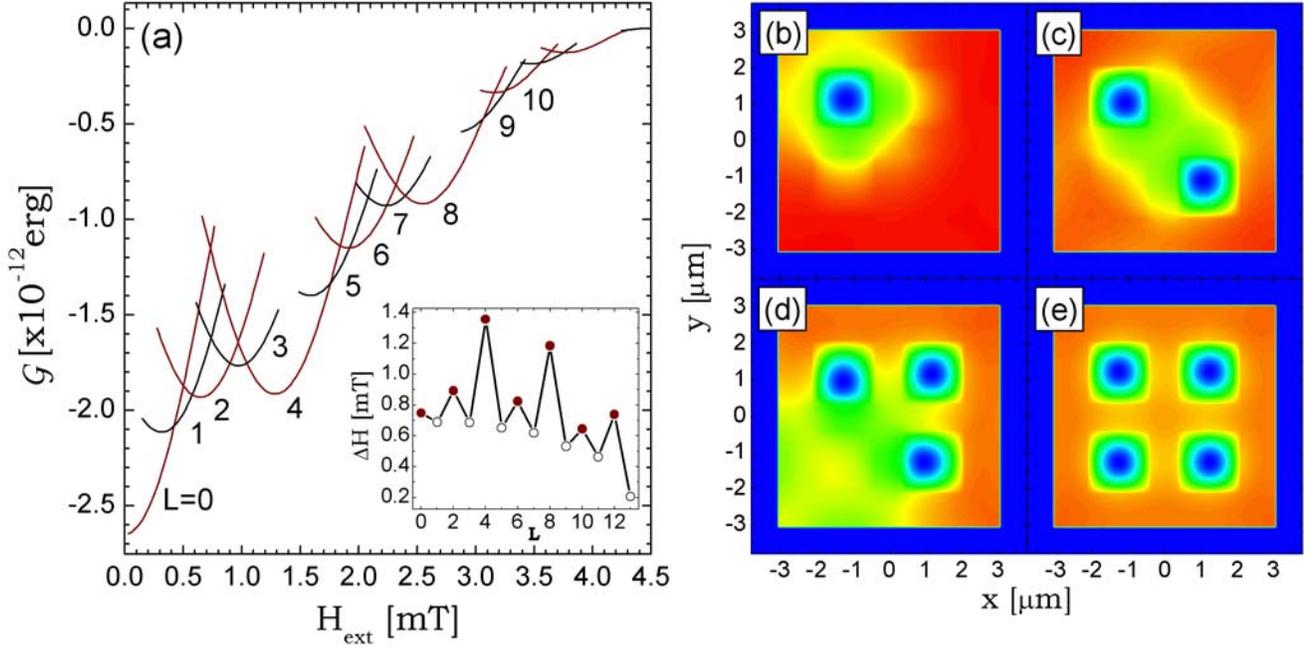

Figure 2 (a) The Gibbs free energy of the system as function of the applied homogenous magnetic field $H_{ext}$. With increasing field, the superconducting ground state transits through different vortex states, denoted by consecutive vorticity numbers (*L*). The inset shows the applied-field-range of stability of each particular vortex state. (b)-(e) The equilibrium position of the vortices, with respect to the position of the blind holes is illustrated by the Cooper-pair density plots (blue/red color – low/high density) for different *L*-states (L=1-4, respectively).

cell logic state remains obscure. In the ground state computation, inputs to a system must excite the system energy and define the new ground state energy for the system (as long as the input is applied). When the input is immediately applied the system state may be hard to predict. But for a defined system with a well-defined input, the state of the lowest energy must be uniquely defined. In order to secure a well-define input for a FCA unit-cell, we propose placing nanoengineered current-carrying circular loops on top of the two diagonal blind holes, as cartooned in Figure 1(a). Our aim is to use the magnetic properties of the current loop, which, when the created magnetic moment is parallel to the external field, attracts the vortices, and vice versa. Using this magnetic pinning property[15,16], one can prepare the FCA input signal at will, by changing the polarity of the applied currents in the loops.

To remain within the experimental verification limits, we present results for the current loops with radius *R=600nm*, and assume them to be separated from the superconductor by a *50nm* oxide layer. This layer is usually used in experiments to prevent the proximity effect between the superconductor and the metallic coils, since they should be only magnetically and not electronically coupled.

To investigate the time evolution of the superconducting state as a response to an external drive (in our case current *I*), we employ discretized time-dependent GL equations[25]. When mapped on a uniform Cartesian grid *(x,y)*, and within an iteration procedure proposed by Schweigert and Peeters in Ref.[20], each iteration step is related to a predefined fraction of the Ginzburg-Landau time, given in the microscopic theory by $\tau_{GL}=4\pi\sigma_n\lambda(T)^2/c^2$, where $\sigma_n$ stands for the normal-state conductivity. Using the above equation, for mesoscopic Al samples we obtained $\tau_{GL}\approx2.8ps$ for considered temperature *T =0.94T$_c$*. The results of our TDGL simulation are presented in Figure 3, as the time relaxation of the ground state, when forced to transit between two degenerate *L=2* states at $H_{ext}=0.645mT$. In Fig. 3(a) the FCA cell is initially in state '1'. In order to change the logic state to '0', we apply the clockwise loop currents *I=0.32mA* which attract vortices (and vice versa in Fig. 3(b)).

As illustrated by Cooper-pair density insets, both transitions occur via the diagonals of the sample. For given parameters, the barrier between the states is lowest on the shortest path connecting the vortices. Over time, they are brought closer together in the center of the sample, and afterwards pushed away from each other to their new equilibrium positions along the other diagonal. One should notice that if currents remain on, the '0' and '1' state do not have the same energy any more (dotted lines in Figure 3), due to the influence of the stray field of the current loops on the superconductor. If for certain purposes it is crucial that '0' and '1' state remain degenerate after the switching process, the current must be turned off after the energy barrier is crossed, as denoted by dashed and solid lines in Figure 3. In such case, the system needs a slightly longer time to equilibrate. Nevertheless, switching time remains under *20ns*, which is unexpectedly fast (if we neglect the finite time necessary for stabilizing the current in the loop $t_{ON}\sim\rho C$). Note that for here discussed parameters, not only samples are rather big, but vortices as well (~0.5μm in diameter), and therefore should be relatively inertial and slow-moving-objects. As will be shown later, there are many possibilities to speed up this process, by more careful design of the cell and/or by an appropriate choice of temperature. In addition, our theory holds for high-$T_c$ superconductors as well, where $\tau_{GL}$ (and consequently switching times) may be >100 times smaller than in aluminum, due to poor conductivity.

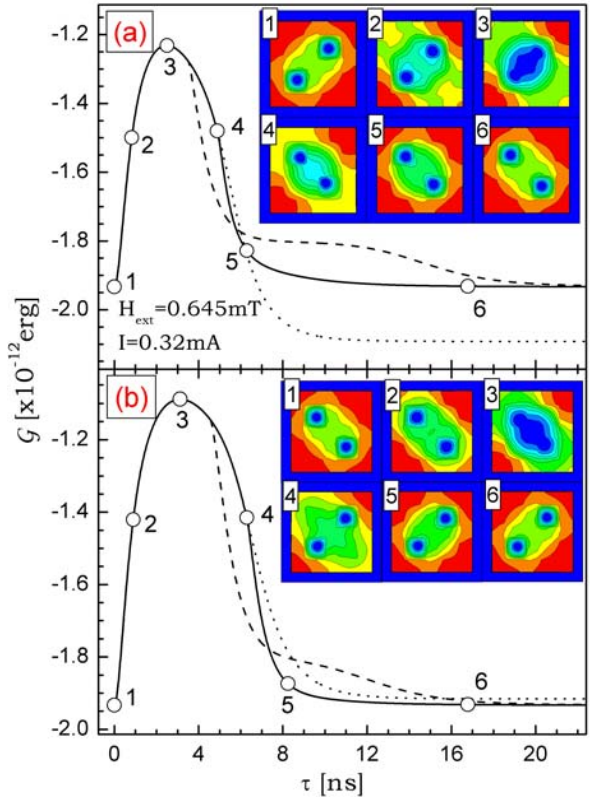

Figure 3 The superconducting free energy during the transition from the '1' state to the '0' state (a) and from the '0' state to the '1' state (b) when a loop current I=0.32mA is applied in the presence of the applied field $H_{ext}$=0.645mT. The insets show snapshots of the transition (contourplots of the Cooper-pair density), taken at intervals indicated by the open circles in the energy curve. For clockwise currents (a), the current was switched off after τ=4.5ns (dashed line) and τ=5.5ns (solid line). The counter-clockwise current (b) was switched off after τ=5.5ns (dashed line) and τ=7ns (solid line). Dotted curves correspond to the case when the current was not switched off.

Of course, the value of the current, necessary for switching between the logic states, may not be chosen freely. In one limit, the applied current could be insufficient to make the vortices overcome the '0'→'1' ('1'→'0') energy barrier. On the other hand, large currents could either over-repel the vortices and push them out of the sample, or become sufficiently strong to self-create an additional vortex in the system. In any case, the logic state would be returnlessly lost. Figure 4 describes these threshold current values as function of the applied magnetic field, in the whole range of stability of the $L=2$ state. It is important to emphasize here the asymmetry in the '0'→'1' with respect to the '1'→'0' transition, also present in Figure 3. Namely, this is a natural symmetry breaking as a consequence of the inverted current drive on the diagonal, while everything else is kept in the same relational order. If certain applications of FCA require absolutely identical transitions to and from a given logic state, two current loops may be alternatively realized on the top 2 blind holes of the input cell, and must contain opposite circulating currents.

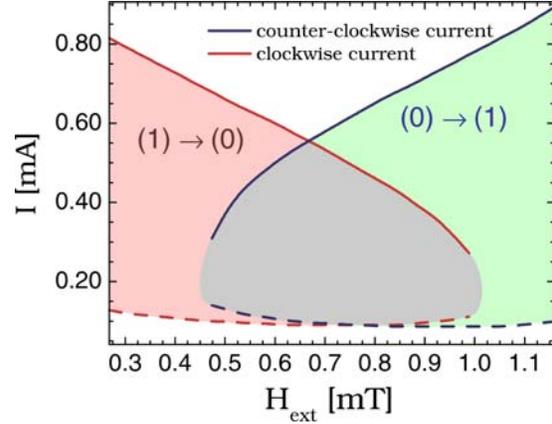

Figure 4 Operating window for the current values in the loops, to induce transitions between the two logic states. The dashed (solid) curves indicate the lower (upper) threshold current. Red curves are related to the '1'→'0' and blue curves to the '0'→'1' transition.

## FCA LOGIC GATES

The FCA unit-cell alone only describes a simple bi-stable device (two possible polarities), but analyzing two or more we begin to see more interesting behavior. In an isolated cell, the two states have identical energy and, hence, are occupied with equal probability. If there are other polarized cells in the immediate vicinity (exposed to the same external magnetic field), the energetically favorite state is determined by repulsion between the vortex fields originating from the neighboring cells. As a result, if one cell is fixed at a particular polarity, a cell next to it will relax to the same polarity. Similarly a linear chain of cells will all assume the same logic state, which is the main principle of the so-called information "wire"[1]. However, very different from the standard QCA case, the interaction between the adjacent cells may be strengthened by conjoining the edges of the cells. Namely, the whole gate may be carved out of one singly connected superconducting sample, in which case the vortices in neighboring cells do not interact solely via fields, but also through the locally induced supercurrents. Of course, careful engineering is needed, to ensure that all cell-parts of such a sample contain exactly 2 vortices. If absolutely necessary, in order to prevent vortices from hopping between the cells, thin superconducting wall can be placed between the cells (an anti-analogue of a blind hole).

Figure 5 shows a 2x2 array of FCA cells, illustrating propagating state from one corner to the opposite one. The signal is first branched into two identical ones (upper and lower cell) before affecting the output cell. The interaction across two corners makes the information transfer more reliable. The 4 involved unit-cells are joined in a single superconducting sample. While keeping the material, thickness of the sample and temperature the same as in previous section, we reduced the lateral dimensions of each cell by 40% (e.g. $w=1.8\mu m$, $w_i=510nm$, $w_0=210nm$, $R=360nm$). This decrease in size speeds up the operation times, and increases the operating value of the external magnetic field. In order to have 2 vortices in each cell in the ground state, we choose the field $H_{ext}=1.373mT$. To further illustrate the rich landscape of technological possibilities, the logic state in the input cell is defined by only one current loop. While the physics involved remains the same, this simplifies the experimental realization of the device.

Initially, due to mutual interaction, all cells are equilibrated in the same logic state (see inset (1) of Figure 5(a)). If we apply a current in the counterclockwise direction to the loop located in the right top corner of the system (white circle in the lower inset) the cell underneath the loop changes its state (insets 2-4) overcoming an energy barrier.

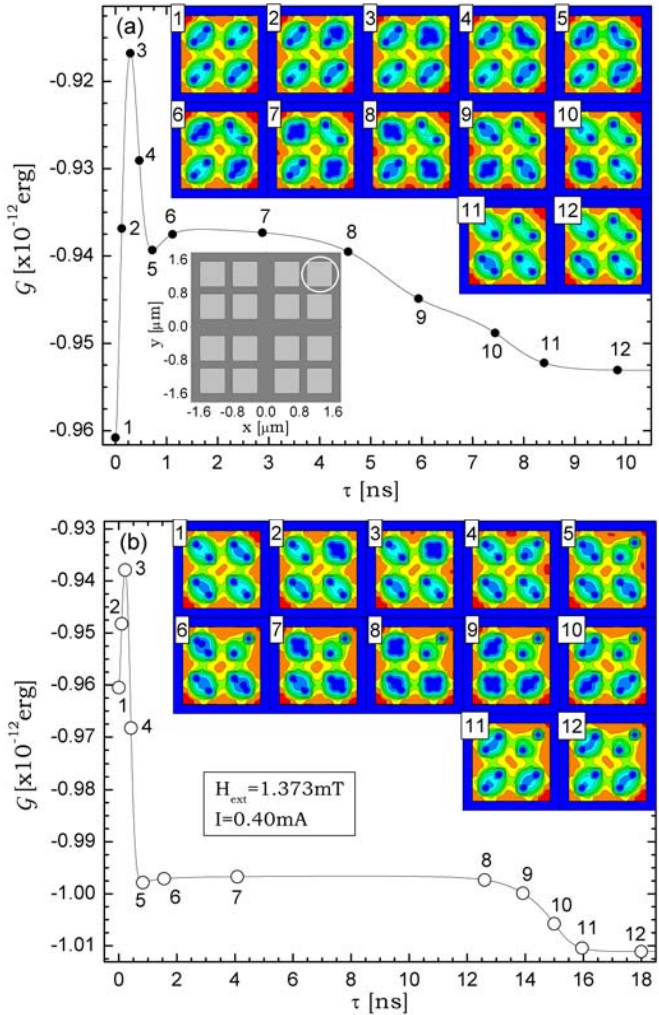

The system then continues to minimize its energy, while the neighboring cells below and left from the input change their logic state (insets 5-9). Finally, vortices in the output cell are triply pushed to the new ground-state position. The process is totally reversible, as the initial state is switched back by changing the current polarity in the loop (Figure 5(b)). As a difference, the '0'→'1' transition takes about 20% longer time, since the current loop does not affect both vortices in the input cell in the same fashion, contrary to the previously described '1'→'0' case. Nevertheless, one should notice that the total operating time is multiply decreased by the 40% reduction of the unit-cell size. This time, including the switching process for 4 cells, is *<20ns* and faster than the single switching demonstrated in Figure 3, even though only one current loop was used (and consequently the time necessary to overcome the energy barrier is larger).

One of the fundamental logic devices is a three-input majority logic gate. In an FCA realization (analogous to QCA), shown in Figure 6, this gate consists of an arrangement of five unit cells: a central logic cell, three inputs labeled A, B, and C, and an output cell. The polarization states of inputs A, B, and C determine that of the central cell, which can assume either polarization, while the output polarization follows that of the central cell. In operation, the polarization of the central cell becomes that of the majority of the three input cells. FCA logic gates can be cascaded, so that in a more complex circuit, the three inputs would be driven by the outputs of the previous gates. Similarly, the output of the majority gate can be connected to drive a subsequent logic gate.

The majority gate circuit actually performs the Boolean function *out=AB+BC+AC*. Therefore, it can be programmed to act as an OR gate or an AND gate by fixing any one of the three inputs. If the programmed input is a '0', the AND operation is performed on the remaining two inputs. If the programmed input is a '1', the OR operation is performed. As an example, we show the operation of the majority gate when having inputs A=0, B=1, and C=1. The input signals are modified in the same fashion as in Figure 5, by one current loop at each input-cell. They are placed in such a way to ensure A=0, B=1, and C=1 for the same value and direction of the current in all three of them: *I=-0.10mA*. The gate is simulated at applied magnetic field of $H_{ext}=1.5mT$ for the size of the samples kept the same as in previous case. In addition to the expectedly well-defined output illustrated by the snapshots of the Cooper-pair density during the transitions (see insets of Figure 6), one should notice that the total operation time, including input formation, is of the order of a nanosecond (i.e. GHz technology).

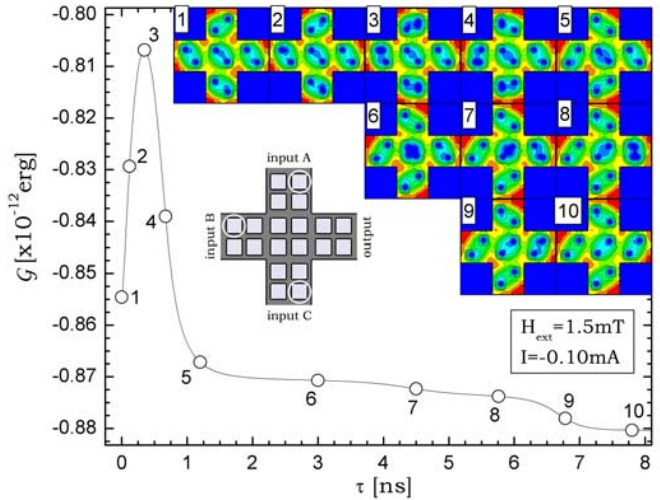

Figure 5 The energy landscape of the information transfer through a 2x2 cell-matrix. The insets depict the Cooper-pair density of the superconducting state during logic switching between the cells, corresponding to the open circles in the free energy curve. The applied field is $H_{ext}=1.373$ and the current in the loop (white circle indicates its position on the sample) is I=0.40mA, for the counterclockwise (a) and clockwise (b) cases. The current remained turned on throughout this ground-state computation, to prevent feedback.

Figure 6 A majority gate. The operation process is illustrated by the Cooper-pair density insets, corresponding to the superconducting states at the open circles in the free energy curve. The applied field is $H_{ext}=1.5mT$ and the current in each of the loops (positioned as indicated by white circles in the lower inset) is I=-0.10mA, defining the input signals as A=0, B=1, and C=1.

## INFLUENCE OF THE PARAMETERS

To summarize, we introduced a new fluxonic concept for Quantum Cellular Automata, where the superconducting micro/nano square with 2x2 blind holes is used as host for 2 vortices, when exposed to homogeneous external magnetic field. For symmetry reasons and their mutual repulsive interaction, those two external flux lines occupy the blind holes along one of the diagonals. Therefore, two degenerate states are possible and may be used to represent logic '0' and '1'. The predefining of the input is realized by adequate placing of current loops on top of the sample, which attract or repel the vortices depending on the current polarity. Two or more FCA unit cells interact with each other magnetically, through the magnetic moments of captured vortices, and/or via the locally induced supercurrents. The possibility of having spatially connected cells enhances the interactions and decreases the computational time, contrary to standard QCA where such realization is not possible.

Although there is no power amplification in FCA, all unit cells in a FCA circuit are experiencing the same conditions (applied field, temperature) and therefore the circuits may have large number of cells without endangering the data transfer. Improving the QCA concept, feedback is easily prevented in FCA by above-mentioned current loops, which can either fix the input signal, or be switched in central cells in order to remember a particular state during the computation. The operating temperatures are also worth mentioning – previously realized QCA based on metal-insulator tunnel junctions or Si quantum dots embedded in a $SiO_2$ matrix, operate at very low temperatures (*~100mK*), as the electrostatic interaction energy must be significantly larger than the thermal one. FCA circuits can operate close to the critical temperature of the material used. Knowing that many well-established $HT_c$ superconductors transit to the normal state at *T~100K* the difference becomes obvious.

In this article, we presented the simulations of immediately experimentally realizable FCA circuits, which are significantly larger than the scale of futuristic electronics. However, our predictions are generally applicable to much smaller superconducting samples. The only condition the device should meet is its relative size of several coherence lengths at given temperature. This characteristic scale makes possible for two vortices to nucleate in the sample, which is crucial for FCA. Nevertheless, even on the micrometer scale, we demonstrated FCA operation frequencies in the gigahertz range, for low-$T_c$ superconductors. Besides going to smaller scales, the operating times can be significantly decreased by carefully tailoring the energy barrier between the '0' and '1' logic state. For example, it is well known that barriers in superconductors diminish when temperatures are closer to $T_c$. To lower the barrier one could also consider operating in a magnetic field range close to the additional flux entering/expulsion, but only if the applied field can be stabilized to a high accuracy (the 2 vortex state may not be disturbed). Optionally, geometrical parameters play a role - shallower blind holes (weaker pinning) or smaller distance between them, facilitate vortex motion from one to another. However, the largest gain can be achieved if FCA is realized in high temperature superconductors. Small characteristic lengths, low barriers and high temperatures ensure extremely low operation times. As a set-back, a problem of thermal fluctuations and bad mechanical properties of these materials must be taken care of. In any case, modern experimental techniques support production of very large number of identifiable superconducting cells, which is crucial for cellular automata. Moreover, various vortex-imaging methods or a very recently developed multiple-small-tunnel-junction method[26] can easily serve as very accessible read-out techniques, bringing out one more advantage of FCA as compared to the quantum-dot automata.

## METHODS

### Numerical approach used

In the description of the properties of thin mesoscopic superconductors in external magnetic field, we rely upon the Ginzburg-Landau (GL) formalism, and solve two coupled nonlinear GL equations self-consistently, following the numerical approach of Schweigert and Peeters[20]. Using dimensionless variables and the London gauge for the vector potential divA=0, we write the system of GL equations for the superconductor with variable thickness $d(x,y)$ in the following form (see Ref. [24] for detailed description), keeping the temperature dependence explicitly:

$$\left(-i\vec{\nabla}-\vec{A}\right)^2\psi = \psi\left[\left(1-\frac{T}{T_C}\right)-|\psi|^2\right] + i\left(-i\vec{\nabla}-\vec{A}\right)\psi\frac{\nabla d(x,y)}{d(x,y)},$$

$$-\frac{\kappa^2}{d(x,y)}\Delta\vec{A} = \frac{1}{2i}\left(\psi^*\vec{\nabla}\psi-\psi\vec{\nabla}\psi^*\right)-|\psi|^2\vec{A}.$$

In order to achieve time relaxation and convergence of our calculation, we add the time derivatives of the order parameter and the vector potential to the left hand sides of the GL equations respectively. This time-dependent (TD) problem is then mapped on a numerical grid using the link variable approach, introduced by Kato *et al.* (see Ref. [25] and references therein). After discretizing TDGL equations, each step in our iterative procedure corresponds to a relaxation time of $\tau_{GL}=4\pi\sigma_n\lambda^2/c^2$ ($\sigma_n$ denotes the normal state conductivity and $\lambda$ the magnetic penetration depth). The results obtained for the order parameter density and the vector potential describe the equilibrium superconducting state, which, in principle, may be metastable. Therefore, after starting the simulations from randomly generated initial conditions, by comparing the Gibbs free energies of all vortex states found, we determine the ground state configuration.

### Acknowledgements
We gratefully acknowledge support from the Flemish Science Foundation (FWO-Vl), the Belgian Science Policy, and University of Antwerp (GOA).
Correspondence and requests for materials should be addressed to F.M.P.

### Competing financial interests
The authors declare that they have no competing financial interest.